\documentclass[prb,aps,twocolumn]{revtex4}
\pdfoutput=1
\usepackage{graphicx}
\usepackage{amsmath}
\usepackage{amssymb}

\begin{document}
\title{Local structure of Liquid-Vapour Interfaces}

\author{Maia Godonoga}
\affiliation{School of Chemistry, University of Bristol, Bristol, BS8 1TS, UK}

\author{Alex Malins}
\affiliation{School of Chemistry, University of Bristol, Bristol, BS8 1TS, UK}
\affiliation{Bristol Centre for Complexity Science, University of Bristol, Bristol, BS8 1TS, UK}

\author{Jens Eggers}
\affiliation{School of Mathematics, University of Bristol, University Walk, Bristol BS8 1TW, UK}

\author{C. Patrick Royall}
\affiliation{School of Chemistry, University of Bristol, Bristol, BS8 1TS, UK}

\date{\today}

\begin{abstract}
The structure of a simple liquid may be characterised in terms
of ground state clusters of small numbers of atoms of
that same liquid.
Here we use this sensitive structural probe to consider
the effect of a liquid-vapour interface upon
the liquid structure. At higher temperatures
(above around half the critical temperature)
we find that the predominant effect of the interface is to
reduce the local density, which significantly suppresses the local
cluster populations. 
At lower temperatures, however, pronounced 
interfacial layering is found. This appears to be connected with
significant orientational ordering of clusters based on 3- and 5-membered rings,
with the rings aligning perpendicular and parallel to the interface respectively.
At all temperatures, we find that the population of five-fold symmetric structures 
is suppressed, rather than enhanced, close to the interface.

\end{abstract}

\maketitle

\section{Introduction}
What is the structure of the transition zone between liquid and vapour? 
In 1893 van der Waals~\cite{rowlinson1979, rowlinson1982} described the 
liquid-vapour interface as a smooth density profile $\rho(z)$, 
shortly preceded by Rayleigh~\cite{rayleigh1892}. 

Early computational~\cite{croxton1971} and theoretical work 
based on the Born-Green-Yvon equation ~\cite{croxton1971ii} suggested
a significant degree of surface layering in the Lennard-Jones model of simple liquids.
However, further simulations~\cite{rao1976} and density functional theory
~\cite{rowlinson1982,evans1992} showed that little surface layering is 
expected for materials based on van der Waals interactions, and that 
the interfacial profile is dominated by a smooth change in density.
Relative to the critical temperature $T_{c}$, metals typically remain liquid
at rather lower temperatures than Lennard-Jones like materials, and here pronounced surface layering is indeed found
\cite{evans1992,chacon2001,tarazona2002,velasco2002,walker2007}.
An important related problem to the free interface is the confinement
introduced by a hard wall, crystal or similar external field, which may be tackled
with density functional theory~\cite{evans1992,lowen1994}, simulation~\cite{heni2000}
and experiment~\cite{huismann1997,yu1999}. In this case, layering
is more generally observed.

In experiments on conventional materials, the measurement of (intrinsic) interfacial
profiles is complicated by collective surface excitations
in the form of capillary waves  
\cite{smoluchowski1908,buff1965,evans1979,stillinger1995}.
For those systems where surface layering is pronounced such as metals, 
the lengthscale above which capillary wave effects
become dominant is around 100 nm. This lengthscale
is larger than that typically encountered in computational studies, 
where capillary waves only cause a small broadening of surface layering~\cite{tarazona2002},
but cannot be neglected in experiment. Nonetheless surface layering has been 
demonstrated in experiments, for example in liquid Mercury~\cite{magnussen1995}. 

A further class of materials which may be directly investigated in real space due to their 
mesoscopic lengthscale are colloidal suspensions. Most notably, colloid-polymer
mixtures exhibit (colloidal) vapour-liquid phase separation~\cite{lekkerkerker1992,poon2002}, moreover
the hard-sphere interactions that closely approximate such systems
are amenable to theory~\cite{brader2003}. Here capillary waves
have been directly visualised~\cite{aarts2004}. So far, to our knowledge, interfaces in colloidal systems
with long-ranged interactions~\cite{elsner2009}
which can be deeply cooled and might exhibit pronounced layering  (the equivalent of metals) have not been studied.
However, single-particle resolution enables other approaches to be used. Notably, a method
introduced by Chacon and Tarazona~\cite{chacon2003} which pins a plane to a layer of surface particles
indeed revealed density oscillations perpendicular to the pinned surface in a colloidal liquid~\cite{royall2007c}.

Here we consider the perturbation induced by the interface on the
liquid structure at the molecular/atomic level. 
We consider two different liquid systems. The first
is a truncated and shifted Lennard-Jones (LJ) liquid~\cite{smit1992,shi2001}
whose triple temperature $T_{tr}^{LJ}$ lies at $T_{tr}^{LJ}/T_{c}^{LJ}=0.630$
where $T_{c}^{LJ}$ is the critical temperature.
To observe a stronger effect of the liquid-gas interface, 
we also consider a model potential which approximates Sodium (Na), 
whose triple point lies at a rather lower temperature relative to criticality of 
$T_{tr}^{Na}/T_{c}^{Na}=0.22$
 \cite{chacon2001,velasco2002}. This model 
is known to exhibit strong surface layering around the triple point, 
which is also seen in \emph{ab initio} simulations \cite{walker2007}.

In the bulk, it is argued that liquid
structure is determined to a significant degree by structures which 
are energetically locally favourable~\cite{frank1952}. Such locally favoured structures 
can correspond to clusters which minimise the potential
energy in isolation and may be catalogued and ordered by the number of particles they contain~\cite{doye1995,wales1997,williams2007,royall2008}. 
Structures which exhibit five-fold symmetry, in particular, are very common
and account for up to 2/3 of all particles~\cite{royall2008,taffs2010}. As a result, the local dynamics 
can be determined to a significant degree by the geometry and 
the symmetries of the locally favoured structures. This observation is 
potentially significant to understand the glassy behaviour of supercooled 
liquids: if the symmetry of the locally favoured structure differs from 
that of the crystalline state, the path to crystallization is frustrated. 

It is thus natural to ask what happens to the concentration of clusters
near the liquid-gas interface. This will be a significant measure of 
local liquid structure. Two mechanisms will contribute to a significant 
change in cluster concentration near the interface. First, 
structures favoured by the bulk may be suppressed 
near the free surface. Second, clusters are non-isotropic, and
might thus be ordered in a particular way with respect to the free surface. 
We will explore this ordering effect in the case of two very common 
clusters, following the nomenclature of Doye \emph{et. al.}~\cite{doye1995}: 
the 5A triangular bipyramid and 7A pentagonal bipyramid, which are 3-and 5-membered rings 
with atoms 
attached above and below the plane of the ring.

One might think that the free surface could lead to an {\it enhancement}
of local fivefold symmetry. The reason is that in the case of deeply
quenched liquids, the free interface can to an extent be thought of
as a constraining field. Now the structure induced in simple liquids
by confinement such as hard walls has been shown, under the application
of a second field, to induce some five-fold symmetry \cite{heni2002},
which may be conjectured to relate to X-ray reflectometry experiments
on lead which found evidence for local fivefold symmetry at the interface
\cite{reichert2000}. Surprisingly, here our analysis suggests the
opposite effect: 7A pentagonal bipyramid clusters, which form the basic 
unit of five-fold symmetry in our methodology~\cite{williams2007,royall2008},
are suppressed disproportionately near the interface. 

To determine locally favoured structures, we use a novel method, the Topological
Cluster Classification (TCC). This identifies small clusters of particles
from within bulk phases which are topologically similar to a set of
reference clusters~\cite{williams2007}. In this case the reference 
structures are formed by groups of $5\leqslant m\leqslant13$ Morse particles
in isolation~\cite{wales1997} and, depending on the range of the Morse 
potential, correspond to the global energy minimum clusters of the 
Lennard-Jones and Sodium potentials considered. We have recently explored 
the structure of the bulk Lennard-Jones liquid and a system similar to 
the Sodium model using this method~\cite{taffs2010}.
Contrary to the suggestion of Frank who conjectured that
in the structure of such liquids 13-membered icosahedral structures
might be prevalent, we found that these only account for one particle in a
thousand, according to our criteria. Instead five-fold symmetry stems
from pentagonal bipyramids, which account for 54\% of the particles at
the Lennard-Jones triple point~\cite{taffs2010}.

This paper is organised as follows. First, we describe our methodology
for identifying the interface, and local structure in section 
\ref{sec:Methodology}. We then proceed to present and discuss our 
findings in section \ref{sec:Results} before placing our findings in 
context in the concluding section \ref{sec:Conclusions}.

\section{Methodology}
\label{sec:Methodology}

\subsection{Models}
We use the Monte Carlo (MC) method 
to simulate a liquid in equilibrium with its vapour phase~\cite{frenkel}.
We use the 
Lennard-Jones (LJ) 12-6 potential, where, for two particles separated 
by a distance $r$, the interaction energy $U$ is given as

\begin{equation}\label{eq:LJ}
\beta U_{LJ}(r)=4\beta \varepsilon_{LJ}\left[\left(\frac{\sigma_{LJ}}{r}\right)^{12}-\left(\frac{\sigma_{LJ}}{r}\right)^{6}\right],\end{equation}
\noindent where $\beta=1/k_{B}T$ where $k_{B}$ is Boltzmann's constant
and T is temperature. There are two parameters: $\varepsilon_{LJ}$ is
the strength of the attraction between the particles and $\sigma_{LJ}$
determines the range of the interaction. From here on we quote all
quantities in standard reduced units, where energies and lengths are
normalized by $\varepsilon_{LJ}$ and $\sigma_{LJ}$ respectively. 

Now the triple point of Lennard-Jones lies around $T_{tr}^{LJ}=0.68$, so
the maximum degree of cooling relative to the critical point is 
$T_{tr}^{LJ}/T_{c}^{LJ}=0.63$. 
We would like to realize even lower temperatures
and this can be achieved by employing models with
longer-ranged interactions which freeze at lower temperatures 
relative to criticality. 
In particular, the spherically symmetric model 
for Sodium (Na) introduced by Chacon \emph{et. al.} 
\cite{chacon2001,velasco2002} has a triple point around $T_{tr}^{Na}=0.27$ 
and critical point around $T_{c}^{Na}=1.25$, enabling a much larger 
degree of cooling, $T_{tr}^{Na}/T_{c}^{Na}=0.22$. This model
potential for Sodium reads

\begin{equation}\label{eq:Na}
\beta U_{Na}(r)= \beta A_{0} e^{-ar}- \beta A_{1} e^{-b(r-R_{1})^{2}},\end{equation}
\noindent
where $A_0=437.96$ eV, $A_1=0.18382$ eV, $a=2.2322$ $\text{\AA}^{-1}$
$b=0.2140$ $\text{\AA}^{-2}$ and $R_{1}=3.5344$ $\text{\AA}$. 
A lengthscale $\sigma_{Na}=3.48$ $\text{\AA}$ is defined as minimum of 
$U_{Na}$ along with a well depth $\varepsilon_{Na}=U_{Na}(\sigma_{Na})=0.1885$
eV. Energies and lengths are then normalized by
$\varepsilon_{Na}$ and $\sigma_{Na}$ respectively.

\begin{figure}
\includegraphics[width=8cm]{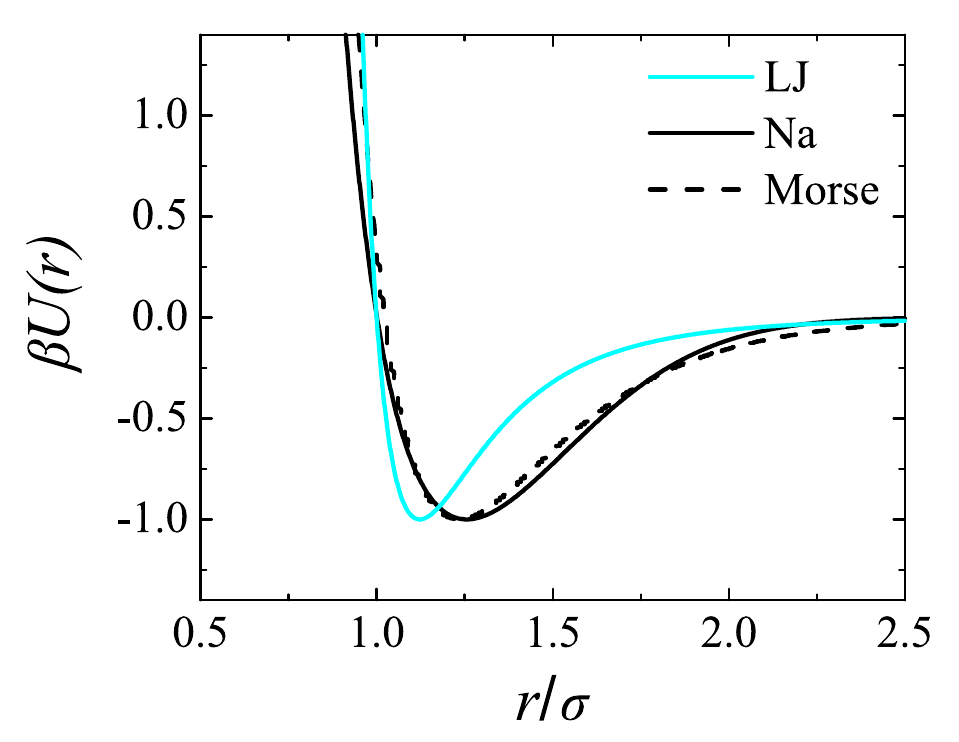}
\caption{\label{figU}Model potentials used. Blue line, Lennard-Jones (LJ); solid black
line, Sodium (Na); dashed line, Morse potential with range parameter
$\rho_{0}=4.05$.}
 \end{figure}

In Fig.~\ref{figU} we show the Lennard-Jones and 
Sodium potentials to be considered below. We also display the Morse potential,
which will be used as a reference potential for the minimum energy clusters of the Sodium model. 

We truncate and shift the Lennard-Jones potential at finite range
$r_{cut}=2.5\sigma$ such that $\beta U_{LJ}(2.5$$\sigma$$)=0$.
In the case of the model Sodium potential, 
we follow \cite{velasco2002} and truncate the model potential at $r_{cut}=2.5\sigma$ 
but, unlike the LJ case, we do not shift the potential,
i.e. $\beta U_{Na}(2.5$$\sigma$$)<0$. 

\subsection{Simulation details} 

We use Monte Carlo simulations in the NVT ensemble~\cite{frenkel}.
The simulations consist of $N=16000$ particles in cuboid boxes with sides $L$x$L$x$2L$
and with periodic boundary conditions. This geometry ensures that, once equilibrated, the liquid forms  
a slab with two interfaces perpendicular to the $z$-direction.
Assuming the interface does not overlap with itself, the $z$-position
of the top or bottom interface may be expressed as a function of
$(x,y)$.

We present results for two state points in the case of Lennard-Jones
$\rho=0.32$, $T=0.95$ and $\rho=0.32$, $T=0.68$, 
and in the case of Sodium $\rho=0.485$, $T=0.542$ and $\rho=0.485$, $T=0.270$. 
These correspond to $T/T_{c}=0.880$, $0.630$, $0.434$ and $0.237$
respectively where $T_{c}$ is the critical point of LJ or Na respectively. 
$\rho$ is the overall density in the simulation box, which contains
both liquid and vapour. This gives $L=29.2\sigma_{LJ}$ for the LJ state points 
and $L=25.5\sigma_{Na}$ for the Sodium state points.

For all simulations suitable liquid-vapour coexistence initial configurations are 
generated by first estimating appropriate bulk liquid $\rho_{L}$ and vapour $\rho_{G}$ densities at coexistence
from the literature \cite{Trokhymchuk1999,chacon2001} and then equilibrating a 
liquid slab of density $\rho_{L}$ in coexistence with a vapour at $\rho_{G}$ for at 
least 300000 MC sweeps. Equilibration is ensured by both checking relaxation of the total potential 
energy and by examining the evolution of the liquid-vapour density profile $\rho (z)$. 
It has been shown that the coexistence densities and $\rho (z)$ depend on the 
system size, the geometry of the box, and the boundary conditions of the simulation \cite{Binder2000}, 
as well as the truncation length and whether the potential is shifted \cite{Trokhymchuk1999}. Once the 
density profile $\rho (z)$ is seen to have stopped evolving, we say the system is at equilibrium and
a production run of 10000 MC sweeps is used to generate 100 independent 
configurations for analysis. Four or more independent simulations are performed for each state 
point and the results averaged over all eight or more liquid-vapour interfaces obtained from these.

\subsection{Determining the interface location}

To calculate the location of the interface we first 
fit $\rho (z)$ with a hyperbolic tangent function in the following form 

\begin{equation}\label{eqtanh}
\rho(z) = \frac{\rho_{L}+\rho_{G}}{2} + \frac{\rho_{L}-\rho_{G}}{2} 
\tanh \left( \frac{z-z_{0}}{w} \right)
\end{equation}

\noindent
where $\rho_{L}$ and $\rho_{G}$ are the bulk densities of the liquid and the gas at 
equilibrium coexistence, $w$ quantifies the width of the interfacial region, and $z_{0}$ 
is the position of the interface with respect to the simulation coordinate
axis $z$. As we will discuss below, $w$ as determined by (\ref{eqtanh})
ignores the effect of capillary waves on the interface. Although it is difficult to accurately 
decouple the intrinsic width of the interface from that of a capillary broadened interface \cite{Binder2000}, 
we show below, for all the state points studied here, ignoring the effects of capillary broadening
and using the fit (\ref{eqtanh}) to determine the position and orientation of the interface, that the 
conclusions drawn about the structure of the interface do not significantly change as a result of this approximation. 

\begin{figure}
\includegraphics[width=8cm]{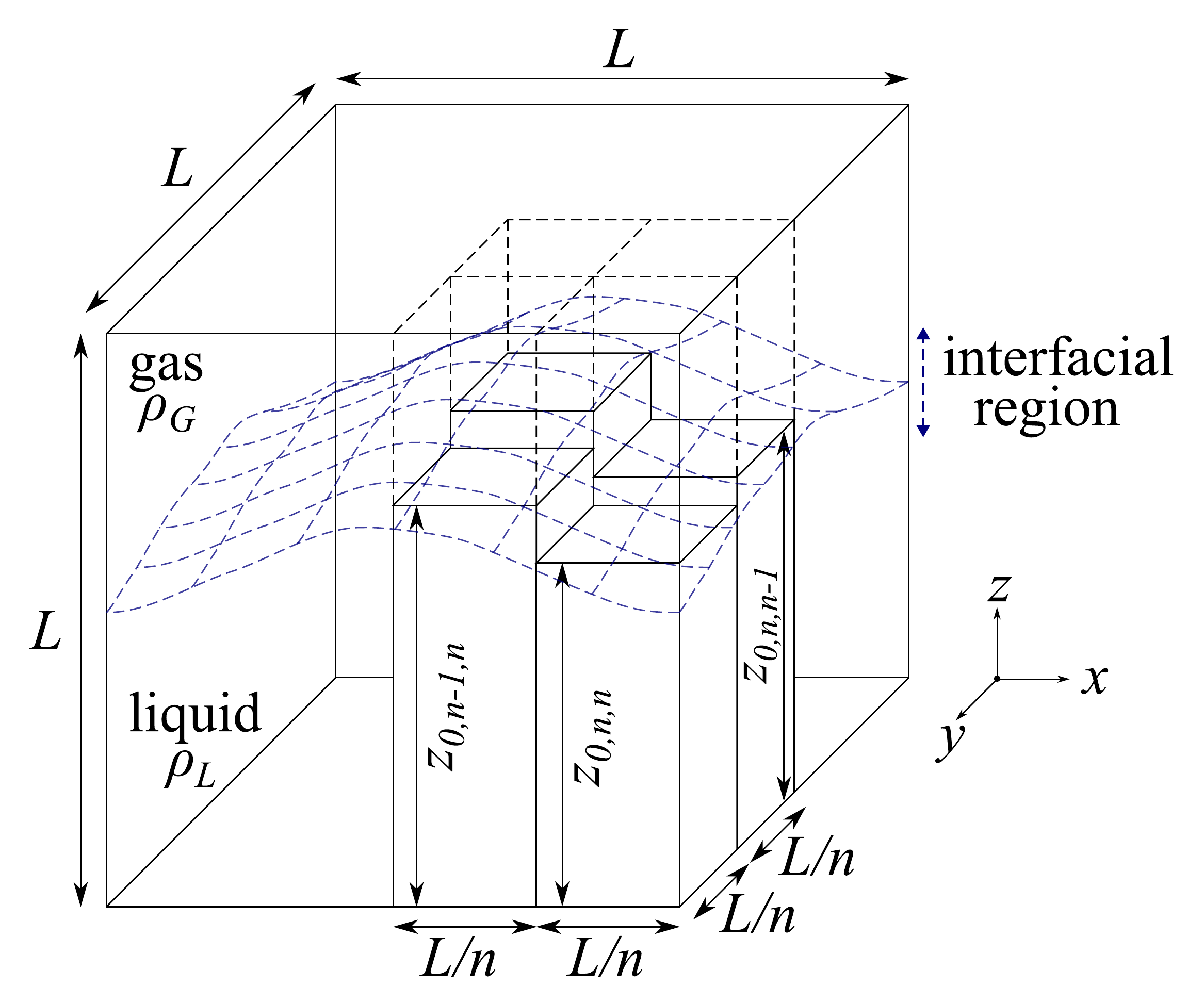}
\caption{\label{figCW}Identification of the position of a capillary perturbed interface. 
The interface is shown in blue and the cube is split into $n^{2}$ columns $i,j$. The interface is 
assumed flat in each column and the position is extracted by a fitting step function (Eq. (\ref{eqstep}))
to the columnar density profile $\rho_{i,j}(z)$.
Note periodic boundaries apply in the $x$ and $y$ directions only.}
\end{figure}

To assess the effect of capillary waves, we note that Eq. (\ref{eqtanh}) 
assumes that the interface position is independent of the local $x,y$ position and 
is perfectly flat, which is clearly not the case if capillary waves are present. We imagine 
that the interfacial width may be decomposed into capillary-broadened and intrinsic
components $w_{cw}$ and $w_{i}$ where $w = w_{cw} + w_{i}$ and we are interested in the intrinsic
width $w_{i}$ \cite{sides1999}. We determine the effect of surface roughening (capillary waves) 
as follows \cite{vink2005}. Firstly we divide the simulation cube into 
$n^{2}$ columns of width and breadth  $\frac{L}{n}$ and height $L$ for integer $n$
as shown in Fig. \ref{figCW}. 
Rather than Eq. (\ref{eqtanh}), we  
model the density profile $\rho_{i,j}(z)$ of column $i,j$ with the step function 

\begin{equation}\label{eqstep}
\rho_{step}(z)=
\begin{cases}
\rho_{L} & z < z_{0,i,j}, \\
\rho_{G} & z \geq z_{0,i,j},
\end{cases}
\end{equation}

\noindent
where $z_{0,i,j}$ is the position of the interface and $i$ and $j$ are
integers denoting the column. 
We minimize the least square residuals between $\rho_{i,j}(z)$ and $\rho_{step}(z)$ 
to find $z_{0,i,j}$.

Re-scaling the $z$ axis as $\rho_{i,j}(z-z_0)$ gives columnar density profiles 
centred on the position of the interface. Taking the mean of these columnar 
profiles gives $\rho_{av}(z-z_0)$ the averaged density profile with capillary 
waves at and above the columnar scale removed, which we then fit with the tanh function (\ref{eqtanh})
leading to a new interfacial width $w_{n}$. Due to capillary broadening, the square of the measured interfacial width
has a logarithmic dependence on the column width \cite{sides1999}, 
while extrapolating the column width to zero gives an estimate of the intrinsic width.
However, for very small column widths, there are insufficient statistics (coordinates) in the columns to 
reliably give the columnar density profiles $\rho_{i,j}(z-z_0)$.

For the LJ $T=0.95$ state point, where we expect the largest contribution $w_{cw}$ from capillary waves, 
the minimum we obtain is 
$w_{n}=2.11(1) \sigma_{LJ}$, compared to the flat-interface value of $w=2.38(2) \sigma_{LJ}$. 
$w$ as obtained from equation (\ref{eqtanh}) is only slightly (around $10\%$) larger than the minimum $w_{n}$ obtained 
from the capillary wave analysis and therefore roughening of the interface is minimal over the lengthscale we measure.
Following this analysis we conclude
capillary waves make a rather limited impact on the interfacial
widths we can measure and hereafter neglect the effect of
capillary waves and simply fit the whole simulation box with Eq. (\ref{eqtanh}) to 
determine the location of the interface. We henceforth rescale our $z$-axis
such that $z_0=0$.

\subsection{Topological cluster classification (TCC)}

\begin{figure}
\includegraphics[width=8cm]{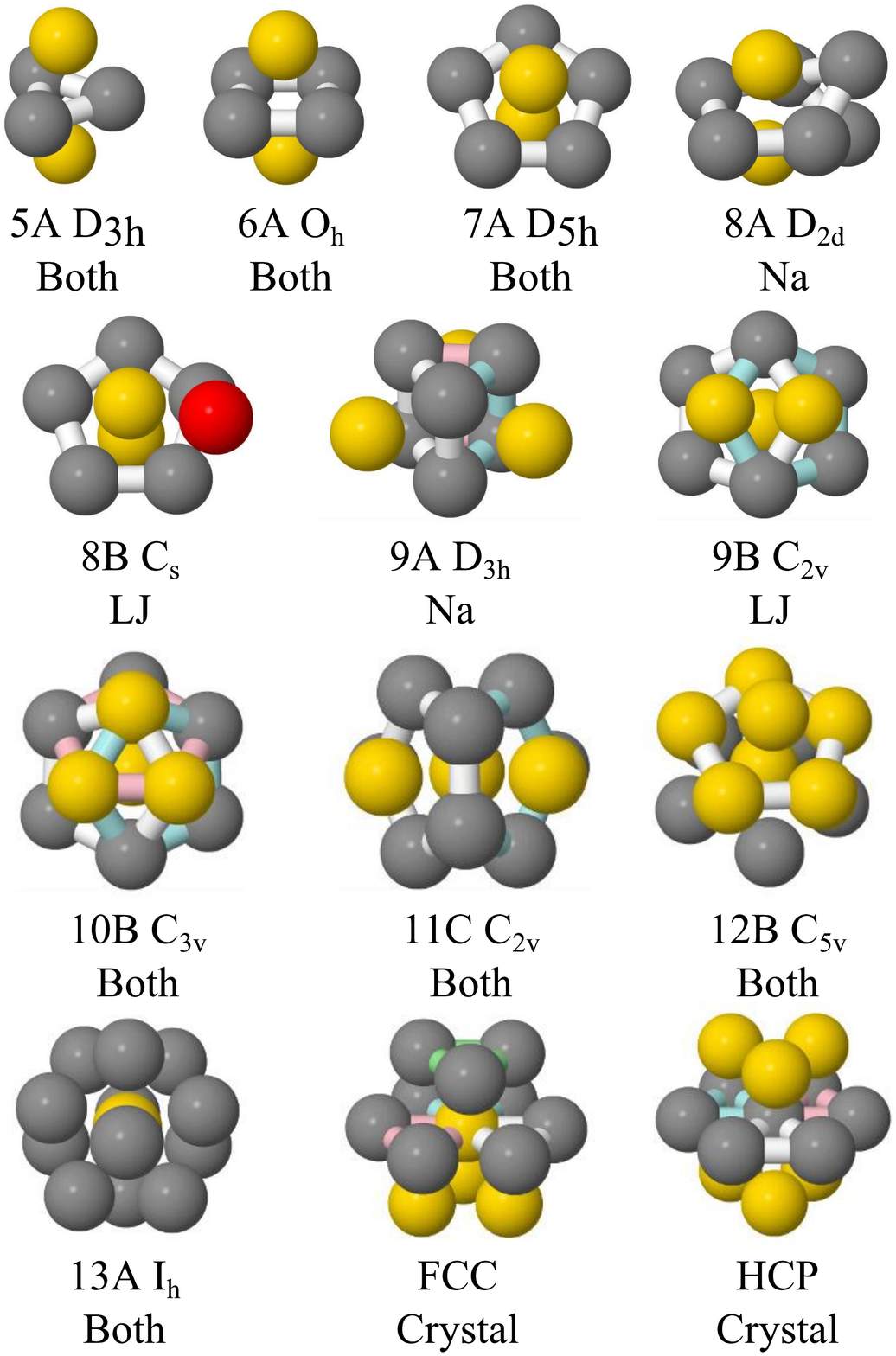}
\caption{\label{figTCC}Structures of LJ and Na clusters, nomenclature following~\cite{doye1995}, and their point group symmetries. 
The colours denote the method used for cluster detection by the TCC algorithm: 3-, 4- and 5- membered 
ring atoms are grey, spindle atoms are yellow, additional atoms to a basic cluster are red (here just 8B which is detected from 7A clusters), 
and rings are coloured white, blue, pink and green.}
\end{figure}

To analyse the structure, we first identify the bond network using
a modified Voronoi construction with a maximum bond length $r_{cut}=1.8\sigma$
and four-membered ring parameter $f_{c}=0.82$ \cite{williams2007}.
Having identified the bond network, we use the topological cluster classification to determine the
nature of the cluster. This analysis identifies all the shortest path
three, four and five membered rings in the bond network. 
These rings form the basic building blocks of our analysis. A 3-membered
ring with atoms bound above and below is identified as a 5A triangular bipyramid,
a 4-membered ring with two bound atoms as a 6A octahedron and a 5-membered
ring as a 7A pentagonal bipyramid, as shown in Fig. \ref{figTCC}.
The additional atoms bonded to the rings are termed spindle atoms. 
Now the ground state clusters for
Lennard-Jones are known \cite{wales1997}, but we are unaware of any work identifying ground state clusters
for the Sodium model employed here. However, ground state clusters for the variable range Morse potential have been calculated  \cite{doye1995}.
We therefore \emph{assume} that the clusters of the Morse potential with the relevant range (as discussed below)
are also ground states for Sodium. We then use the
TCC to find clusters which are global energy minima of the Lennard-Jones
and Morse potentials, applying the latter to our Sodium results. Supporting this assumption is 
work which shows that for a model for Sodium with many-body interactions, the Gupta model, 
the ground state clusters are the same as those we infer for the Chacon and Tarazona model \cite{Lai2002,Noya2007}.

The Morse potential is defined

\begin{equation}\label{eqMorse}
\beta U_{M}(r)=\beta\varepsilon_{M}e^{\rho_{0}(\sigma-r)}(e^{\rho_{0}(\sigma-r)}-2),\end{equation}

\noindent where $\rho_{0}$ is a range parameter and $\varepsilon_{M}$
is the potential well depth. 
The Morse potential has a variable range and we choose the range such that it closely approximates the Sodium model. 
The extended law of corresponding
states \cite{noro2000} provides a means by which different systems
may be compared with one another, by equating their reduced second
virial coefficients

\begin{equation}\label{eqReducedVirial}
B_{2}^{*}=\frac{B_{2}}{\frac{2}{3}\pi\sigma_{EFF}^{3}},\end{equation}

\noindent where $\sigma_{EFF}$ is the effective hard sphere diameter
and the second virial coefficient

\begin{equation}\label{eqCorrespondingStates}
B_{2}=2\pi\intop_{0}^{\infty}drr^{2}\left[1-\exp\left(-\beta U(r)\right)\right].\end{equation}

\noindent The effective hard sphere diameter is defined as

\begin{equation}\label{eqEffective}
\sigma_{EFF}=\intop_{0}^{\infty}dr\left[1-\exp\left(-\beta U_{REP}(r)\right)\right].\end{equation}

Varying the Morse range parameter such that $B_{2}^{*}$ for Sodium
and the Morse potential are equal gives $\rho_{0}=4.05$. Now the
ground state minima for the Morse potential are known \cite{doye1995} and
here we assume those for $\rho_{0}=4.05$ also hold for $U_{Na}(r)$.
The potentials are compared in Fig. \ref{figU}.
We identify all topologically distinct Morse ($\rho_{0}=4.05$) and
Lennard-Jones clusters~\cite{wales1997}. In addition, we identify the
FCC and HCP thirteen particle structures in terms of a central particle
and its twelve nearest neighbours. We illustrate these clusters in
Fig. \ref{figTCC}. In the case of the Morse potential, for $m>7$
there is more than one cluster which forms the ground state, depending
on the range of the interaction~\cite{doye1995}. We therefore consider
ground state clusters for each system and for $m<14$. For more details
see~\cite{williams2007}.

\section{Results}
\label{sec:Results}

\begin{figure}
\includegraphics[width=8cm]{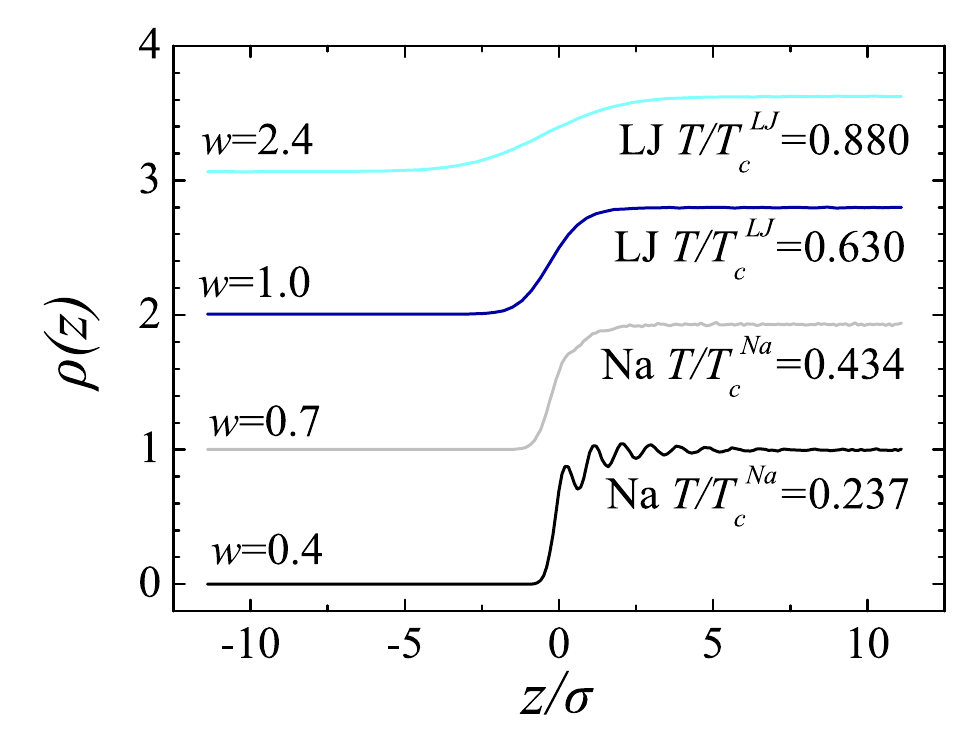}
\caption{\label{figProfiles}Interfacial profiles $\rho(z)$.
From
top, Lennard-Jones $T=0.95$, $T=0.68$, Sodium $T=0.542$, $T=0.297$. 
Data offset for clarity. The sodium interface at the lowest temperature shows
a very pronounced surface layering. The interfaces have all been centred at $z=0$.}
\end{figure}

We begin our presentation of the results by showing the interfacial
profiles of the different systems and state points. In Fig. \ref{figProfiles}
we see that the interfacial width drops with decreasing $T/T_c$. The lowest
temperatures, and thus the smallest interfacial widths, are realized for the 
sodium system. We note the pronounced interfacial layering at the lowest 
temperature \cite{chacon2001,velasco2002}. This indicates that the effect
of the interface on the local ordering of the liquid extends far into the 
bulk in that case. 

\begin{figure*}
\begin{center}
\includegraphics[width=16cm]{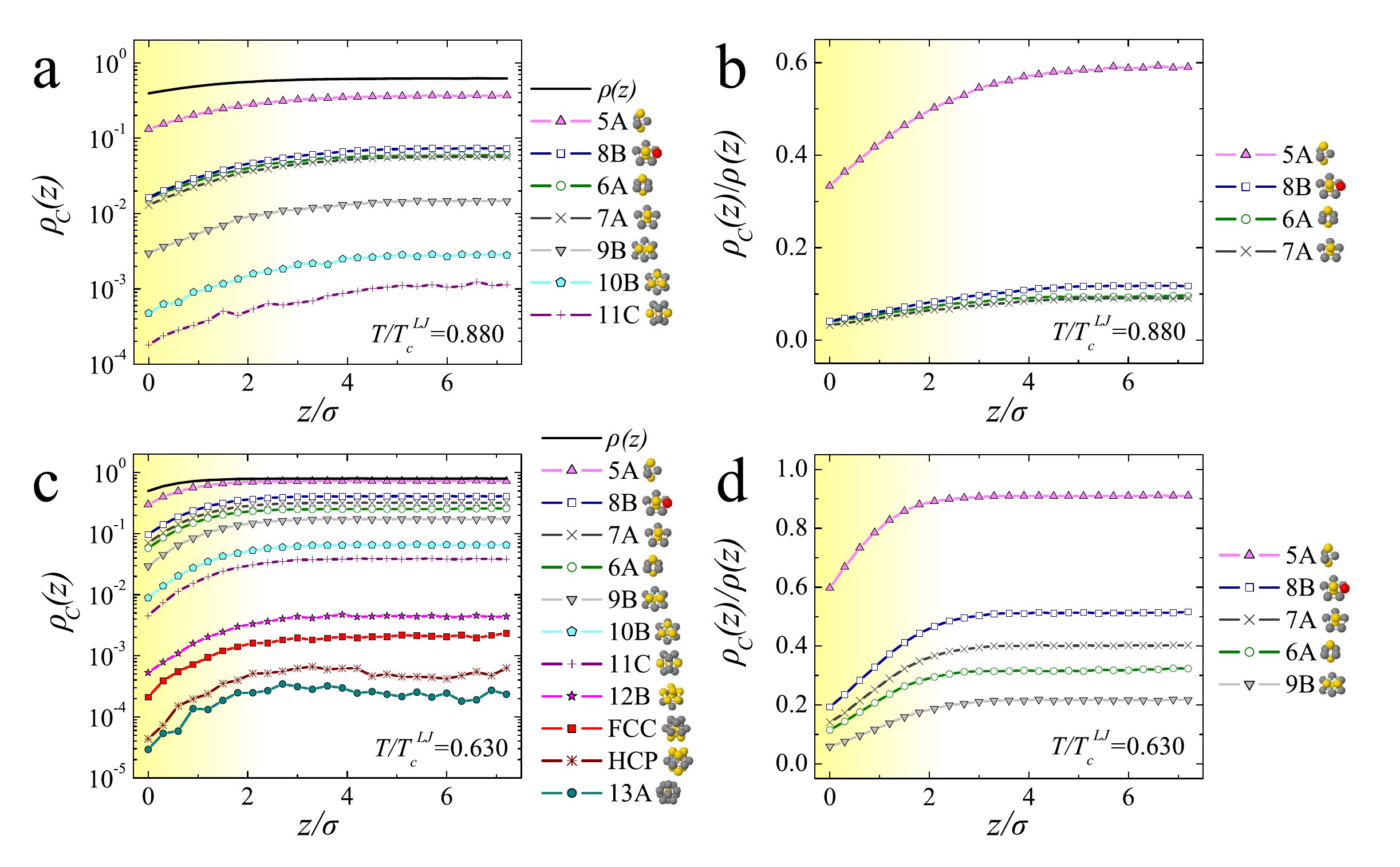}
\caption{\label{figLennardJones}TCC analysis of the Lennard-Jones interface
$\rho_{C}(z)$ denotes the density of particles belonging to a particular
cluster type. $\rho(z)$ reproduces the data in Fig. \ref{figProfiles}.
(a) $T=0.95$ ($T/T_{c}^{LJ}=0.880$). (c) The triple point $T_{tr}^{LJ}=0.68$
($T/T_{c}^{LJ}=0.630$). 
(d) Same data as (c) but normalised by $\rho(z)$.
Interface increases from $z=0$, black interface density
profile.  Yellow shading is a guide to the eye, indicating the width of interfacial region.}
\end{center}
\end{figure*}

We now proceed to the TCC analysis of the Lennard-Jones liquid at two different 
temperatures. We plot the total density of particles $\rho(z)$ as well as the 
particle densities $\rho_C(z)$ for the most popular clusters. These 
are defined as the densities of all the particles belonging to a
particular cluster. 

The densities are plotted on a logarithmic scale as a function of 
the distance $z$ from the interface in Fig. \ref{figLennardJones}. 
Note that the liquid side
only is plotted ($z>0$), and that, for Lennard-Jones, the density
varies smoothly across the interface for our treatment 
\cite{chacon2001,velasco2002}. For each of the temperatures, we also 
plot the cluster densities {\it relative} to the total density on a linear scale. 
The relative density is then defined as the density of particles in a given
cluster $\rho_{C}(z)$ divided by the total density $\rho(z)$.
In the bulk it is seen that the most popular clusters, 
including the 7A pentagonal bipyramid, account for more than half of all the particles. 

As one moves towards the interface, the relative density of clusters 
decreases. For the Lennard-Jones system, our analysis suggests that there is no enhancement of
5-fold symmetry at this interface. In fact, as all cluster populations
drop with the reducing density at the interface, we argue that there
is more of a reduction in 5-fold symmetry than anything else. In particular
the 7A pentagonal bipyramid cluster may be taken as a rough proxy
for five-membered rings, the basic unit of 5-fold symmetry. These
clearly follow the density profile, although the increase with depth
into the liquid is much more marked than the overall density. Similar
conclusions can be drawn about the 13A icosahedron.
The decrease in relative cluster density can be rationalized by observing 
that while potential energy considerations enhance cluster
populations~\cite{taffs2010}, so too does packing~\cite{williams2007}.
Thus we expect a decrease in cluster population at lower density, for
example upon approach to the interface.

For example, for both $T=0.95$ ($T/T_{c}^{LJ}=0.880$) and 
$T_{tr}^{LJ}=0.68$ ($T/T_{c}^{LJ}=0.630$), the relative decrease in  
density amounts to a factor of two in the case of 7A pentagonal bipyramid clusters. 
Since the change appears coupled to the overall density $\rho$, we see  
further change in the structure
of the liquid near the surface for the temperatures studied. 
Note that although it would be attractive to investigate the
effect of reducing the density on the cluster population in a bulk system
of course this is not possible because the Lennard-Jones system
would be unstable to vapour-liquid phase coexistence.

To gain more insight into the effect of the interface on the cluster population, 
it would therefore be desirable to make the 
effect of the interface more pronounced by going to lower temperatures,
thus making the interface sharper [for which we use the model Sodium potential Eq. (\ref{eq:Na})]. 

Note that the triple point
in general has a rather higher population of clusters than does $T=0.95$,
which is not unreasonable, as in addition to the higher density
the enhanced relative attractions at
reduced temperature would be expected to promote clustering, which
has been observed in the bulk \cite{taffs2010liquids}. 

\begin{figure*}
\begin{center}
\includegraphics[width=16cm]{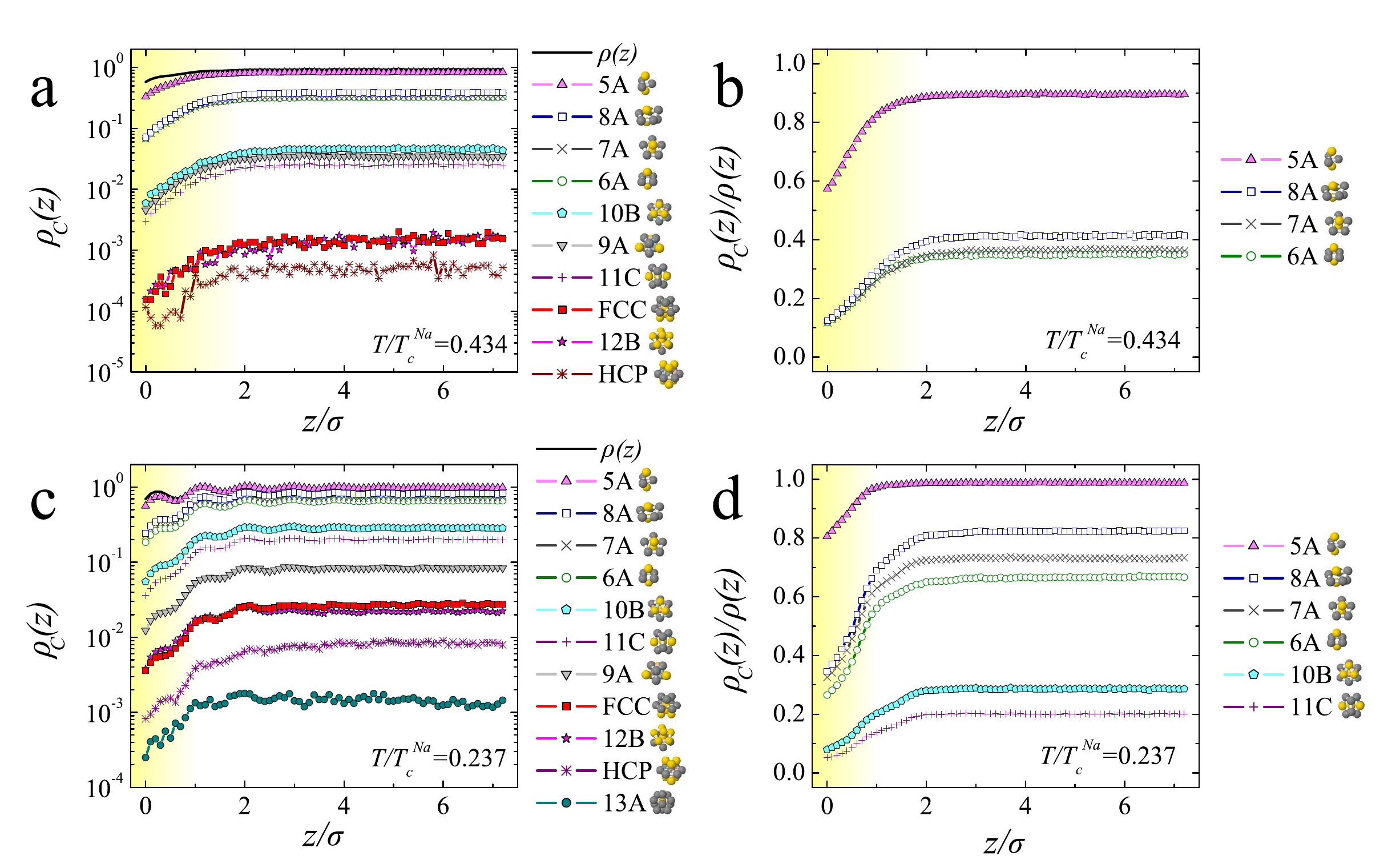}
\caption{\label{figSodium}TCC analysis of the Sodium interface $\rho_{C}(z)$
denotes the density of particles belonging to a particular cluster
type. $\rho(z)$ reproduces the data in Fig. \ref{figProfiles}. 
(a) $T=0.542$ ($T/T_{c}^{Na}=0.434$). (c) Near the triple point $T_{tr}^{Na}=0.297$
($T/T_{c}^{Na}=0.237$). 
(b,d) Same data as (a,c) but normalised by $\rho(z)$.
Yellow shading is a guide to the eye and indicates the interfacial region.}
\end{center}
\end{figure*}

We next performed the same spatially resolved cluster 
analysis for Sodium at lower temperatures relative to criticality. 
We confirmed that, at higher $T_{tr}^{Na}=0.63$, the results with Sodium
were very similar to those for Lennard-Jones.
As Fig.~\ref{figProfiles}
has shown, the interface becomes narrower and at the lowest temperature there are
oscillations of the bulk density profile indicating significant surface layering. 
This shows that the interface changes structuring into the liquid at longer 
ranges than in the case of Lennard-Jones. 

Once more we first show the absolute densities for all the clusters
on a logarithmic scale, and then the densities relative to the total on a linear scale (Fig.~\ref{figSodium}). 
The latter are a direct measure of the influence of the interface on the 
local ordering. At the higher temperature, the decrease in the relative
cluster density is more pronounced than before, but still remains
modest. We note that once more there is certainly no {\it increase}
of 7A pentagonal bipyramid cluster concentration owing to orientational ordering near 
the interface. 

At the lower temperature, 7A clusters account for 82\% of the particles in 
the bulk, the relative density is reduced to 31\% near the interface, a somewhat
larger drop in population.
However, in stark contrast to the higher temperature state points,
the effect on cluster concentration extends far beyond
the density decrease related to the interface, and into the bulk. This mirrors 
the layering as reflected in the oscillations of the density 
profile. The cluster concentrations are evidently a sensitive measure
of the layering effect, and of the change in orientational order it 
introduces. 

\begin{figure}
\includegraphics[width=8cm]{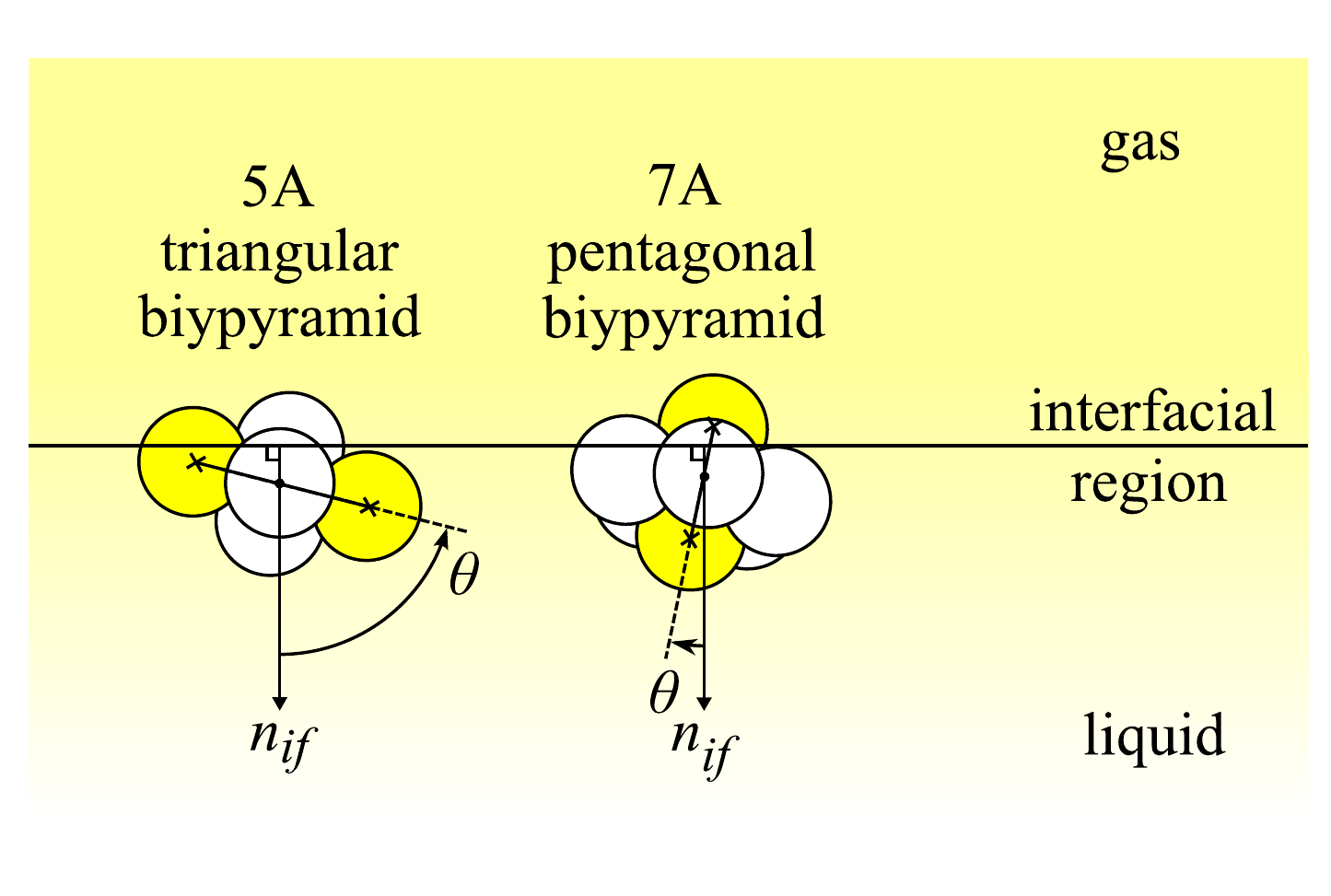}
\caption{\label{figOrientation}Definition of the orientational order parameter used in the case of 
the 5A triangular bipyramid and 7A pentagonal bipyramid. $\theta$ is the angle between 
the normal $n_{if}$ to the plane of the interface (marked in blue) and the vector 
connecting the centres of the yellow spindle atoms.}
\end{figure}

By contrast, the relative decrease of the triangular bipyramidal 5A clusters 
is relatively mild. As the clusters are smaller, this might be expected,
since the perturbation by the interface on the compact 5A structure
will be smaller. 
Other popular clusters, notably 6A octahedra and 8A $D_{2d}$
also show a similar behaviour.
To gain more insight into the concentration of 5A and 7A clusters,
we consider their orientational ordering near the interface. 
As seen in Figs.~\ref{figTCC} and~\ref{figOrientation}, both are viewed as a 3-membered and 
5-membered ring, respectively, with two spindle particles sticking 
out in a direction perpendicular (ignoring thermal fluctuations) to the ring. We neglect the 6A
octahedron due to its high symmetry.
To measure the orientation
of these structures relative to the interface, we consider the angle 
$\theta$ between the spindle particles and the normal to the plane. 
From this we can construct an order parameter
$1/2<3\cos^2\theta-1>$
which is zero in the bulk, where all orientations are equal, unity
in the case that all clusters are aligned with the axis
perpendicular to the interface and $-1/2$ when clusters
lie with the axis parallel to the interface.

\begin{figure}
\includegraphics[width=8cm]{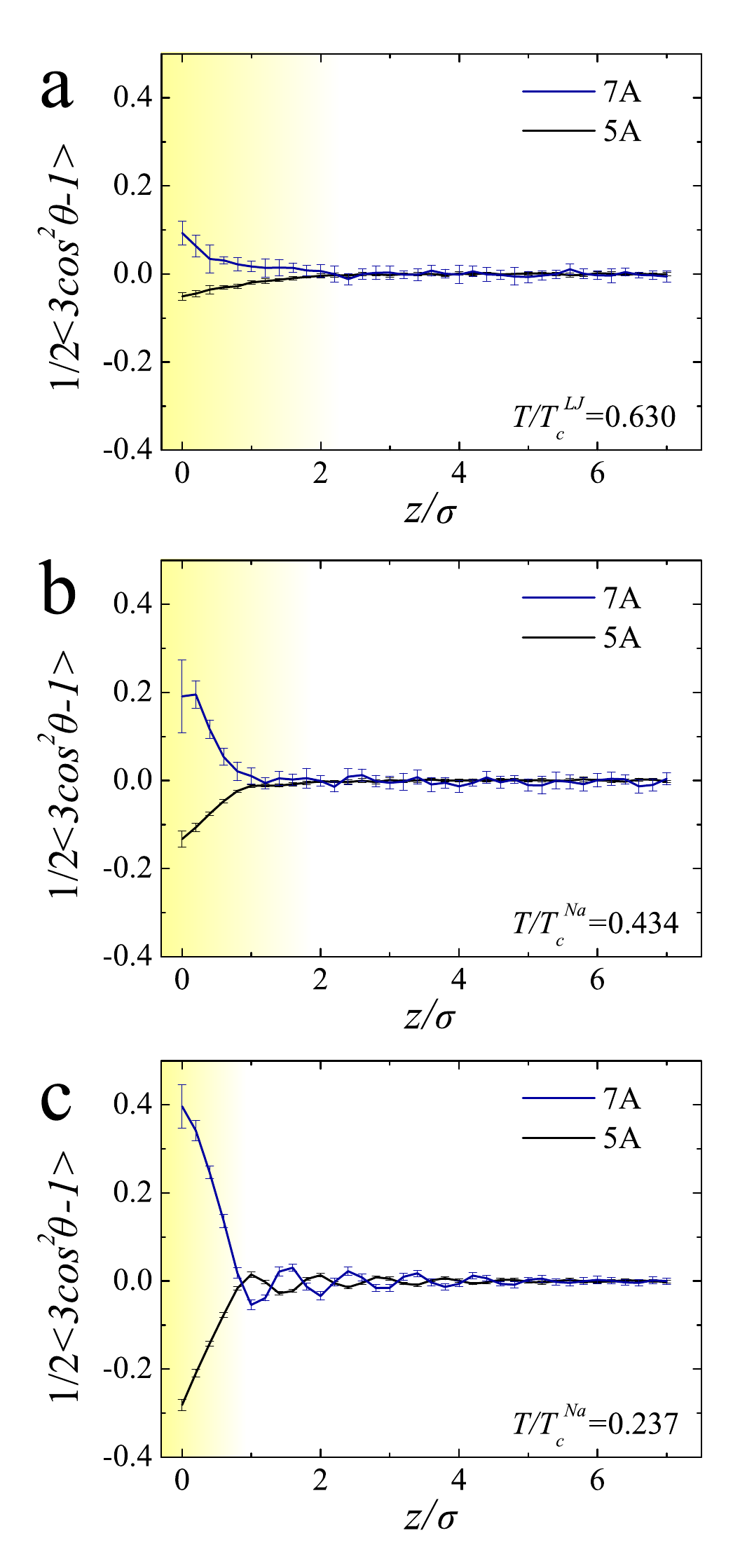}
\caption{\label{figNematic}The ordering of the 5A and 7A clusters 
relative to interface, as function of the distance $z$ from the 
interface. The angle $\theta$ is the angle between the vector that connects
the spindle particles and the normal to the surface. 
(a) Lennard-Jones, $T=0.95,T/T^c=0.880$; (b) Sodium, $T=0.542,T/T^c=0.434$; 
(c) Sodium, $T=0.297,T/T^c=0.237$.}
\end{figure}

In Fig.~\ref{figNematic} we plot the order parameter as function of $z$
for both 5A and 7A clusters for three different $T/T_{c}$ ratios. 
Remarkably, the two types of clusters 
behave in an opposite way near the interface. While the 7A clusters
are aligned with the five-membered ring parallel to the interface, the 5A 
clusters are aligned with its three membered ring {\it perpendicular}
to the interface. This effect increases strongly with decreasing temperature,
so at that lowest temperature [Fig. \ref{figNematic}(c)] in the case of 7A the angle $\theta$ is reduced, 
and the order parameter goes to 1/2, half of what
would be a perfectly ordered state. 

Thus it is likely that the 5-membered ring of the 7A clusters tends to be found
in the plane of the interface. Assuming that 7A is in a configuration which minimises
bond stretching/compression (i.e. zero-temperature potential energy minimum), 
this would leave one end of the spindle 
sticking out from the free surface. This would be an energetically unfavourable
situation, which could explain the suppression of 7A clusters near
the interface. However the consequences of finite temperature may well play an important
role here, a point to which we return in the next section.
The situation is very different for 5A clusters, whose
degree of ordering is also rather less pronounced. More crucially, though
the ordering is such that the spindle remains in the fluid, while the
three-membered ring is oriented parallel to the interface. This permits
the cluster to remain inside the fluid relatively undistorted, and the
effect on its concentration remains small. 
Fig. \ref{figNematic}(c) also shows that the surface layering is 
well reflected by the orientational ordering of the clusters. 

\section{Conclusions}
\label{sec:Conclusions}
We have shown that our topological cluster classification
provides insight into the local liquid structure close to a free interface, and
can directly probe for local structures with fivefold symmetry.
The analysis we have performed provides no evidence in support of an enhanced five-fold symmetry 
near the interface, however those five-fold symmetric structures that are present tend to be aligned
with five-membered rings parallel to the interface.
 
At higher temperatures, the predominant effect appears to be related to the lowering in density 
induced by the interface. This is reflected by a drop in the populations of all the clusters
we consider. Upon reducing temperature, the interfacial density profile exhibits layering, and 
seems to induce a change in cluster population extending well into the dense liquid, long
after the mean density is that of the bulk liquid. Thus we argue that our analysis indicates that 
five-fold symmetry is \emph{suppressed} at the liquid-vapour interface in the systems we have considered.

We use our analysis to reveal orientational information on two basic clusters. Close to the interface, 
the five-fold symmetric 7A pentagonal bipyramid is oriented  with its five-membered ring lying parallel to the interface, conversely
the 5A triangular bipyramid is oriented with its 3-membered ring perpendicular to the interface. Both these effects are
strongly enhanced at lower temperature.

Our analysis is topological in nature. That is to say, the structures identified are based on bonds, and we do 
not consider structural distortions. In particular, we note that the 7A pentagonal bipyramid, aligned with a surface,
might leave a spindle atom exposed. However, for a small amount of strain, this exposed spindle could in fact lie
very close to the plane of the five-membered ring. We suspect that it may do so at finite temperature. 

We identify two possible extensions of this work. It would be interesting to consider the effect of a wall or
similar external field, as an alternative to the free interface. It would also be interesting to 
investigate the behaviour of clusters other than those which are minimum energy ground states in isolation, 
for example 5-membered rings. We have seen that the population of 7A pentagonal bipyramids is reduced near the interface, the topological basis of our analysis precludes ruling out that an exposed spindle atom might be related to this  suppression, although at finite temperature we expect this effect to be small. An extension of our analysis to consider distortions in the clusters identified would be helpful. Both these possibilities will be investigated in the future.

\section*{Acknowledgement(s)} Bob Evans is thanked for helpful discussions,  
Stephen Williams for developing the original implementation of the TCC code, and John 
Russo for critical reading of the manuscript.
Hajime Tanaka is thanked for kindly providing space in his lab for CPR and AM while some
of this work was carried out.
CPR thanks the Royal Society for
funding, AM acknowledges the support of EPSRC grant EP/5011214. This
work was carried out using the computational facilities of the Advanced
Computing Research Centre, University of Bristol - http://www.bris.ac.uk/acrc/.

\end{document}